\documentclass[superscriptaddress,reprint,graphicx]{revtex4-1}

\usepackage{graphicx}

\begin{document}

\title{Femtosecond photoelectron and photoion spectrometer with vacuum ultraviolet probe pulses}

\author{Markus Koch}
\email[]{markus.koch@tugraz.at}
\affiliation{Stanford PULSE Institute, SLAC National Accelerator Laboratory, Menlo Park, CA 94025, USA}
\affiliation{Institute of Experimental Physics, Graz University of Technology, Petersgasse 16, A-8010 Graz, Austria}

\author{Thomas J. A. Wolf}
\affiliation{Stanford PULSE Institute, SLAC National Accelerator Laboratory, Menlo Park, CA 94025, USA}

\author{Jakob Grilj}
\affiliation{Stanford PULSE Institute, SLAC National Accelerator Laboratory, Menlo Park, CA 94025, USA}
\affiliation{Laboratoire de Spectroscopie Ultrarapide, Ecole Polytechnique F\'{e}d\'{e}rale de Lausanne EPFL, 1015, Switzerland}

\author{Emily Sistrunk}
\affiliation{Stanford PULSE Institute, SLAC National Accelerator Laboratory, Menlo Park, CA 94025, USA}

\author{Markus G\"{u}hr}
\affiliation{Stanford PULSE Institute, SLAC National Accelerator Laboratory, Menlo Park, CA 94025, USA}

\date{\today}

%\pacs{} 

\begin{abstract}
We describe a setup to study ultrafast dynamics in gas-phase molecules using time-resolved photoelectron and photoion spectroscopy. 
The vacuum ultraviolet (VUV) probe pulses are generated via strong field high-order harmonic generation from infrared femtosecond laser pulses. 
The band pass characteristic in transmission of thin indium (In) metal foil is exploited to isolate the $9^{\text{th}}$ harmonic of the 800\,nm fundamental (H9, 14\,eV, 89\,nm) from all other high harmonics.
The $9^{\text{th}}$ harmonic is obtained with high conversion efficiencies and has sufficient photon energy to access the complete set of valence electron levels in most molecules. The setup also allows for direct comparison of VUV single-photon probe with 800\,nm multi-photon probe without influencing the delay of excitation and probe pulse or the beam geometry. 
We use a magnetic bottle spectrometer with high collection efficiency for electrons, serving at the same time as a time of flight spectrometer for ions.
Characterization measurements on Xe reveal the spectral width of H9 to be $190\pm60$\,meV and a photon flux of $\sim1\cdot10^{7}$\,photons/pulse after spectral filtering.
As a first application, we investigate the S$_1$ excitation of perylene using time-resolved ion spectra obtained with multi-photon probing and time-resolved electron spectra from VUV single-photon probing.
The time resolution extracted from cross-correlation measurements is $65\pm10$\,fs for both probing schemes and the pulse duration of H9 is found to be $35\pm8$\,fs.

\end{abstract}

%\pacs{}
\maketitle

%%%%%%%%%%%%%%%%%%%%%%%%%%%%%%%%%%%%%%%%%%%%%%%%%%%%%%%%%%%%%%%%%%%%%%%%%%%%%%%%%%%%%%%%%%%%%%%%%%%%%%%%%%%%%%%%%%%%%%%%%%%%%%%%%%%%
%%%%%%%%%%%%%%%%%%%%%%%%%%%%%%%%%%%%%%%%%%%%%%%%%%%%%%%%%%%%%%%%%%%%%%%%%%%%%%%%%%%%%%%%%%%%%%%%%%%%%%%%%%%%%%%%%%%%%%%%%%%%%%%%%%%%
\section{Introduction}

Time-resolved photoelectron spectroscopy (TRPES) is a unique method to investigate photoinduced ultrafast relaxation processes in molecules~\cite{Wu2011,Stolow2008,Hertel2006,Stolow2004}. 
As in other ultrafast methods, molecular dynamics in covalent states are induced by the interaction with an excitation pulse typically in the visible or ultraviolet range. The excited state molecular wavepacket is projected onto different electronic continua by an ionization pulse.
The energy flow in a molecule after photoexcitation can be followed in time as transient changees in the intensity and energy of photoelectron (PE) peaks.
In order to access the excited state dynamics, the probe photon energy has to exceed the excited state ionization potential. 
TRPES has allowed the deduction of ultrafast photoexcited dynamics of many molecules ranging from simple diatomic systems~\cite{Vrakking1996}  to polyenes~\cite{Blanchet2001,Blanchet1999} and more complex clusters~\cite{Stolow2004,Verlet2005,Lehr1999}. For the polyenes, TRPES has allowed the separation of nuclear relaxation from electronic relaxation by applying Koopmans' single electron propensity rules~\cite{Blanchet2001,Blanchet1999}.
Probe pulses are generally obtained by crystal based nonlinear optics, limiting the probe pulse photon energy to $<6$\,eV. Those probe photon energies barely ionize the excited state, and nuclear relaxation due to intramolecular vibrational redistribution (IVR) and geometrical relaxation results in an immediate decrease of the signal due to a loss of Franck-Condon overlap. Electronic relaxation due to non-adiabatic transitions~\cite{Domcke2011,Levine2007,Domcke2004,Yarkony1996} will also result in a decay of the photoelectron signal due to a) changing dipole matrix elements and b) further nuclear relaxation in a different electronic state. In general, it is difficult to separate nuclear from electronic relaxation with low photon energy probe pulses, since both processes lead to similar decay patterns. In addition, this method cannot follow further molecular dynamics after the initial decay processes, since the ability for probing is lost.
Strong field multi-photon ionization provides the opportunity to ionize from excited states and from the ground state. However, its interpretation requires extensive strong field ionization calculations~\cite{Matsika2013} and may be further complicated by transient shifts of intermediate resonances.

Vacuum ultraviolet (VUV) probe photons, on the other side, have large enough photon energies to follow relaxation of the excited non-equilibrium state all the way to the vibrationally hot ground state. 
Paired with the short pulse characteristics obtained with high-order harmonic generation (HHG), they are advantageous as probe pulses in TRPES. 
A difficulty that arises with probing in the vacuum ultraviolet and extreme ultaviolet range is a persistent background PE signal from the (unexcited) electronic ground state. This requires a high signal to noise ratio to isolate excited state dynamics.
We use HHG, a tabletop generation mechanism for VUV pulses based on strong field interaction of near infrared femtosecond laser pulses with noble gas atoms. At intensities of $\sim10^{14}$\,W/cm$^2$ part of the valence electrons are ionized from the atoms and accelerated in the laser field. Upon reversal of the electric field the electrons can recombine with the remaining ion and their kinetic energy is converted into photon energy~\cite{Krause1992,Kulander1993,Schafer1993,Corkum1993}, resulting in the emission of a comb of odd numbered harmonics of the fundamental laser frequency. 
Femtosecond VUV pulses can also be obtained from free electron lasers at high fluences but also high costs~\cite{Andruszkow2000,Ackermann2007,Allaria2012}. 

To perform PE spectroscopy, a single harmonic is usually selected from the HHG spectrum by monochromizing beamlines using gratings~\cite{Grazioli2014,Frietsch2013,Igarashi2012,Wernet2011,Nugent-Glandorf2002,Dakovski2010} or multilayer mirrors~\cite{Bunermann2012,Tilborg2009,Gagnon2008,Mathias2007,Siffalovic2001}. Grating monochromators even allow a bandwidth reduction below the inherent width of the respective harmonic at the cost of photon flux~\cite{Frietsch2013,Wernet2011}. 
Pulse broadening in the time domain due to pulse front tilting of diffracted pulses can be corrected with time compensated dual grating setups~\cite{Igarashi2012} preserving the pulse duration of high-order harmonics which is much shorter than that of the driving pulse~\cite{Christov1997}. However, these setups reduce the beamline transmission and increase the alignment effort of the setup. 
%and the best overall pump--probe time resolution documented so far with a dual grating monochromator is 125\,fs with the main contribution of 115\,fs coming from the VUV pulse duration~\cite{Frietsch2013}.
Multilayer mirrors are tailored to reflect a single harmonic. Other harmonics are reflected as well, albeit with reduced efficiency, due to a finite bandwidth and finite specular reflectivity of the materials used. 
A combination of two multilayer mirrors allows to reduce the reflection of undesired harmonics to as low as $\sim1$\%~\cite{Gagnon2008}. 
VUV pulse durations around 10\,fs and below achieved with both grating~\cite{Igarashi2012} and multilayer~\cite{Allison2010,Gagnon2008} monochromatization have been reported.

Here we present a very simple and cost efficient way to select a single harmonic as pump or probe pulse for TRPES with VUV light. We exploit the bandpass structure in transmission of a 150\,nm thick indium (In) metal filter which is centered at harmonic nine (H9, 89\,nm, 14\,eV) of a 800\,nm fundamental. 
In combination with a focusing optic (MgF$_2$ protected Al, or unprotected Au mirror) the transmission contrast is $20:1$ to higher harmonics and $>70:1$ to lower harmonics. We expect negligible temporal broadening of the VUV pulse, as it is the case with multi layer mirrors~\cite{Gagnon2008} or time-delay compensated monochromators~\cite{Igarashi2012}. While the photon energy of H9 is sufficient to ionize many molecules from the ground state it has also the advantage of a higher conversion efficiency in the HHG process because it is closer to the perturbative region~\cite{McPherson1987}.

VUV light was used in gas phase TRPES as a probe pulse to study molecular dynamics of valence electronic states such as the dissociation of Br$_2$ upon 400\,nm excitation to the lowest dissociative state~\cite{Nugent-Glandorf2001,Nugent-Glandorf2002a,Strasser2007,Wernet2009}. 
The VUV light can also be used to excite a system to Rydberg states which are transiently ionized by visible or UV pulses~\cite{Bunermann2012,Kornilov2010,Haber2009a,Kosma2008,Strasser2006,Johansson2003,Cacciani2001,Ubachs2001,Sorensen2000}. Time-resolved  dynamical studies in cationic states, for example in ethylene, have also become possible with VUV pump excitation and IR~\cite{Tilborg2009} or VUV~\cite{Allison2010} probe.

In the following we describe the pump-probe setup, in particular how harmonic 9 is spectrally isolated from the harmonic spectrum (Experimental Setup). The probe path of our setup allows for changing from VUV to the 800\,nm fundamental without alteration of the pump-probe delay and without need for realignment, simply by removing the In filter and deactivating HHG.
This is useful for alignment purposes but, more importantly, also enables multi-photon ionization as a complementary ionization mechanism.
Direct comparison of VUV single-photon probing with multi-photon 800\,nm ionization is thus possible. 
We use a time-of-flight (TOF) spectrometer to record mass spectra for multi-photon probe which is also described in the Experimental Setup. For H9 single-photon probe we detect the electron kinetic energy and operate the TOF spectrometer as magnetic bottle spectrometer to benefit from the higher collection efficiency. We present calibration measurements with Xe gas to estimate the spectral purity, width and photon flux of H9. 
Finally, in the Results and Discussions section we present time-resolved pump-probe studies on the fluorescence dye perylene. For 400\,nm pump excitation we compare 800\,nm multi-photon probe with H9 single-photon probe and determine the spectrometer time resolution for both schemes.

%%%%%%%%%%%%%%%%%%%%%%%%%%%%%%%%%%%%%%%%%%%%%%%%%%%%%%%%%%%%%%%%%%%%%%%%%%%%%%%%%%%%%%%%%%%%%%%%%%%%%%%%%%%%%%%%%%%%%%%%%%%%%%%%%%%%
%%%%%%%%%%%%%%%%%%%%%%%%%%%%%%%%%%%%%%%%%%%%%%%%%%%%%%%%%%%%%%%%%%%%%%%%%%%%%%%%%%%%%%%%%%%%%%%%%%%%%%%%%%%%%%%%%%%%%%%%%%%%%%%%%%%%
\section{Experimental Setup}

\begin{figure}[h]
\centering
\includegraphics[width=85 mm]{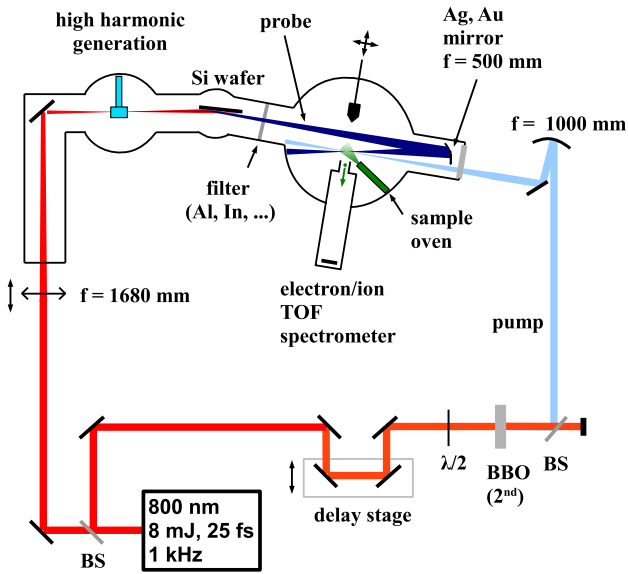}
\caption{Setup for time-resolved photoelectron/photoion spectroscopy based on HHG. In the probe path high-order harmonics are generated in a gas cell. Reflection from a Si wafer at grazing incidence removes the major part of the 800\,nm fundamental and H9 is spectrally isolated by a combination of an In filter and a MgF$_2$ coated Al mirror. Multiple harmonics can be used as probe by using an Al filter and 800\,nm multi-photon probing is possible without filter and deactivated HHG.
In the pump path the fundamental is frequency upconverted and a pump-probe time delay is introduced. The sample is evaporated into the interaction region by an oven. A TOF spectrometer is used to detect electrons in a magnetic bottle configuration or ions by applying a positive repeller voltage. }
\label{fig:setup}
\end{figure}

Fig.~\ref{fig:setup} shows a scheme of the VUV time-resolved photoelectron/photoion spectrometer. This beamline was added to a previously existing setup for time-resolved VUV and soft x-ray (SXR) spectroscopy, which has recently been characterized~\cite{Grilj2014}. 
An amplified laser system (Mantis oscillator and Legend Elite Duo amplifier from Coherent) generates 800\,nm, 8\,mJ, 25\,fs pulses at a 1\,kHz repetition rate, up to 4.5\,mJ of which are available for HHG to obtain VUV probe pulses. The rest of the 800\,nm fundamental can be frequency doubled or tripled for UV pump excitation. A time delay between pump and probe pulses is introduced by a translation stage in the pump arm.

%%%%%%%%%%%%%%%%%%%%%%%%%%%%%%%%%%%%%%%%%%%%%%%%%%%%%%%%%%%%%%%%%%%%%%%%%%%%%%%%%%%%%%%%%%%%%%%%%%%%%%%%%%%%%%%%%%%%%%%%%%%%%%%%%%%%
\subsection{Optical setup}

\begin{figure}[h]
\centering
\includegraphics[width=85 mm]{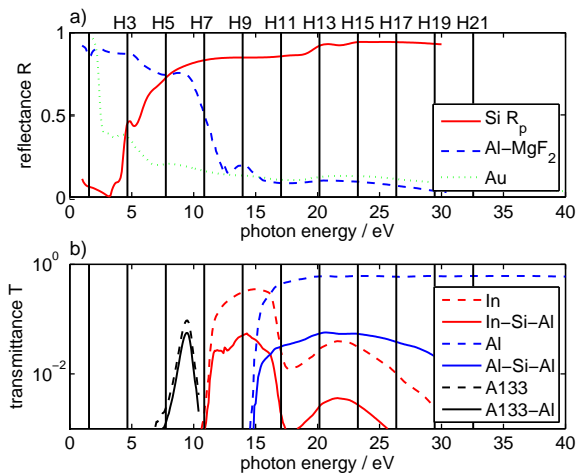}
\caption{(a) Reflectance $R$ of the reflective elements used in the setup. For silicon the angle of incidence is 81 degrees and the reflectance is calculated for p-polarized ($R_p$) light using the Fresnel equations and data from Ref.~\cite{Palik1998}. The reflectance curves of the MgF$_2$ coated Al and the unprotected Au mirror are obtained from McPherson Inc.~\cite{privatecommunication2012}
(b) Transmittances $T$ are shown for 150\,nm thick In and Al filters (dashed lines)~\cite{Henke1993}. The total transmittance function of the beam line (solid lines) is obtained as combination of each filter with the the reflectivity of the Si wafer and Al mirror. In addition, the transmittance of a bandpass filter (Acton Optics A 133) is shown.
Vertical lines indicate the energies of the odd harmonics.
}
\label{fig:optics}
\end{figure}

To generate high harmonics, we focus the 800\,nm pulses with a $f=1680$\,mm lens into a gas cell (27\,mm length, 4\,mm inner diameter) inside a vacuum chamber. The cell contains argon gas at a pressure of about 10\,torr. 
The laser perforates the copper foil seals of the cell, which are replaced every couple of days.  
The fundamental (p-polarized) and high-order harmonics co-propagate after the gas cell and are reflected from a silicon wafer (purchased from University Wafer) in order to reduce a major amount of the fundamental~\cite{Kojima2012,Takahashi2004}. The angle of incidence is chosen to be 81 degrees as a compromise between reduction of the 800\,nm light ($\sim7$\%) and good reflection of the harmonics ($\sim85$\% for H9), as shown in Fig.~\ref{fig:optics}a. This reduction of the fundamental is crucial to prevent destruction of the metal filters that follow in the beam path.

To select the $9^{\text{th}}$ harmonic (H9, 14\,eV, 89\,nm) we use a combination of an In filter and an Al mirror. Fig.~\ref{fig:optics} shows reflectance (a) and transmittance curves (b) of elements that are useful for the selection of a variety of harmonics. Additional information for all elements is listed in Tab.~\ref{tab:optics}.
A MgF$_2$ coated Al mirror with a radius of curvature of 1000\,mm ($f=500$\,mm) focuses the diverging beam into the ionization region of the TOF spectrometer. The Al mirror can be replaced by an unprotected gold mirror for better reflectance at higher photon energies.
Ultra-thin In foil is characterized by a bandpass structure centered at H9 (dashed red curve in Fig.~\ref{fig:optics}b).  In combination with the Si wafer and the Al mirror the total filter curve (solid red curve) spectrally isolates H9 with an estimated transmittance of ~5\%. The transmission contrast on the blue side to H11, H13, H15 is $\sim20$:1 and on the red side to H7 it is $>70$:1. This asymmetry is advantageous as lower harmonics are generated with higher efficiency. These transmittance and contrast values are similar to beamlines using two multilayer mirrors~\cite{Gagnon2008}.
Replacing the In with an Al filter results in a transmittance of multiple harmonics from H11 upwards, with the upper limit depending on the actual cut-off of the Al (or Au) mirror.
Transmittances of the filters is lowered by a few per cent because of a nickle mesh support (optional for Al).
We also note that the Al filter might be covered by an Al$_2$O$_3$ layer due to oxidation, in which case the transmittance is reduced by 30-60\%~\cite{Grilj2014}.
Although not used for this work we prepare for the isolation of H6 at 9.3\,eV.
%Harmonic 15 can be isolated with a Si-Sc multilayer mirror (lower image, green curve) with a relatively high $T$ of 23\%. In this case the multilayer mirror is responsible for focusing and dispersion, as it has been realized in previous experiments~\cite{Siffalovic2001,Mathias2007}. 
Even harmonics can be generated by using a two color field consisting of the fundamental and the second harmonic~\cite{Kim2005} and H6 can be isolated with a bandpass filter (A133, Fig.~\ref{fig:optics}, black lines).

The fundamental is blocked to $2\cdot10^{-7}$ by the In filter ($3\cdot10^{-9}$ with the Al filter)~\cite{Palik1998}.
After separation from their fundamental the diverging harmonic(s) traverse the spectrometer chamber before they are focused by the Al (or Au) mirror into the ionization region of the time of flight (TOF) spectrometer (Fig.~\ref{fig:setup}). To avoid background signal due to ionization by the incoming beam the ionization region and the TOF entrance is located slightly above the incoming beam. 

\begin{table*}[t]
	\centering
	
\begin{tabular}{cccc}
\hline 
optical element & product \# & notes & company\tabularnewline
\hline 
\hline 
Si mirror &  &  & University Wafer\tabularnewline
\hline 
Al mirror & 1200-CSM & MgF$_2$ coating, f = 500 mm & Acton Optics \& Coatings, Princeton Instruments\tabularnewline
\hline 
Au mirror & AU-B-MPC-1025-1000 & f = 500 mm & Lattice Electro Optics Inc.\tabularnewline
%\hline 
%Si-Sc mirror &  & multilayer & Frauenhofer IOF, Jena, Germany\tabularnewline
\hline 
In filter &  & 150 nm thick, Ni mesh support & Lebow, Luxel\tabularnewline
\hline 
Al filter &  & 150 nm thick, Ni mesh support & Lebow, Luxel\tabularnewline
\hline 
A133 & 133-XN-1D & 2.5 mm MgF$_2$ substrate & Acton Optics \& Coatings, Princeton Instruments\tabularnewline
\hline 
\end{tabular}

\caption{Optical elements used in the beamline to spectrally isolate and deliver harmonics to the electron/ion spectrometer.}
	\label{tab:optics}
\end{table*}

Al metal filters are routinely used in HHG setups to eliminate the remaining fundamental radiation, as they transmit harmonics in the 17 to 70\,eV range (see Fig.~\ref{fig:optics}). Sn metal filters exhibit a transmission window between 16 and 24\,eV which allows selection of harmonics 11 to 15. This bandpass structure was exploited in one arm of a split mirror interferometer to realize a femtosecond pump--probe experiment with a pair of VUV pulses (the second arm used a multilayer mirror)~\cite{Allison2010}.
In another setup, a resonance energy that was included in the relatively broad multiband reflectivity of a multilayer mirror was optionally eliminated by the sharp Sn transmission edge at 24\,eV~\cite{Bunermann2012}.
Semiconductor (Si) and metallic (Al, Zr) filters have also been used for amplitude and phase shaping of high-order harmonics aiming at the pulse duration  reduction of attosecond pulses~\cite{Gustafsson2007}.

Our setup allows for optimization of the HHG conversion efficiency, i.e. phase matching of the HHG process, by measuring the flux of photoelectrons  emitted from a Au surface with a low noise charge sensitive amplifier (SRS model SR 570)~\cite{Grilj2014}. 
In the TRPES setup the Au surface can be brought into the beam after the Si wafer and Al filter (optional) but in front of the In filter.
HHG is optimized by adjusting the focus position with respect to the cell, the cell gas pressure and the chirp of the fundamental laser pulse. However, we note that pure optimization for maximum HHG power might result in increased spectral width of the harmonics~\cite{Heyl2011,Farrell2009}. Instead, optimization on the width and strength of a photoelectron reference spectrum (e.g., that of Xe, see below) resulted in better spectral quality of the HHG source.

The pump beam is branched off from the laser output by a 20\% beamsplitter (BS, Fig.~\ref{fig:setup}) and traverses a delay stage for pump-probe delay adjustment. The 800\,nm fundamental is then frequency doubled in a BBO crystal and the 400\,nm light is separated by a dichroic beam splitter. A halfwave plate in front of the BBO crystal allows to adjust the pump power from zero to a few hundred microjoules. A $f=1000$\,mm Al mirror is used to focus the pump beam into the ionization region through a 3\,mm thick CaF$_2$ window. Spatial overlap of pump and probe beam can be optimized with the help of a phosphor screen that can be moved into the interaction region and viewed with a CCD camera. To find time overlap both beams can be guided to a photo diode.

%%%%%%%%%%%%%%%%%%%%%%%%%%%%%%%%%%%%%%%%%%%%%%%%%%%%%%%%%%%%%%%%%%%%%%%%%%%%%%%%%%%%%%%%%%%%%%%%%%%%%%%%%%%%%%%%%%%%%%%%%%%%%%%%%%%%
\subsection{Focus size}

We measure the excitation and probe beam diameter at the ionization region with a phosphor screen that can be brought into the beam and viewed with a camera. 
The full width at half maximum (FWHM) of the beam diameter is determined as average of two Gaussian fits to the horizontal and vertical intensity distribution. For the 800\,nm fundamental we find a beam diameter of $\sim90$\,$\mu$m.
To guarantee good spatial overlap we use a bigger diameter of the 400\,nm pump beam of 250\,$\mu$m.
We determine the diameter of the harmonics using the Al filter and MgF$_2$ coated Al mirror (the HHG spectrum consists primarily of H11 to H15, see below) and find a diameter of $\sim94$\,$\mu$m.
This diameter is the demagnified image of the HHG source region. With a (de-)magnification of $M=-0.18$ of our beamline we estimate a HHG source diameter of about 520\,$\mu$m, which is consistent with the harmonic generation geometry.

%%%%%%%%%%%%%%%%%%%%%%%%%%%%%%%%%%%%%%%%%%%%%%%%%%%%%%%%%%%%%%%%%%%%%%%%%%%%%%%%%%%%%%%%%%%%%%%%%%%%%%%%%%%%%%%%%%%%%%%%%%%%%%%%%%%%
\subsection{The sample}

A molecular beam of perylene (Sigma Aldrich, white powder, purity $\geq99.5$\%) is introduced into the TOF ionization region by a home-built capillary-oven~\cite{McFarland2013a}.
Due to a capillary of 0.9\,mm inner diameter and $\sim40$\,mm length the oven is able to produce a much narrower beam than a conventional Knudsen cell. Depending on the required sample density of the spectroscopic scheme the oven is heated to temperatures in the range of $100^{\circ}$C to $170^{\circ}$C, corresponding to perylene vapor pressures of $10^{-4}$ to $10^{-1}$\,torr~\cite{Inokuchi1961}. Based on deposition measurements on a quartz micro-balance we estimate a sample density at the exit of the capillary ranging from $10^{11}$ to $10^{14}$\,cm$^{-3}$ for the given temperature range~\cite{McFarland2013a}.
As the cone angle of the molecular beam increases significantly from $\sim10^{\circ}$ (FWHM) at lower vapor pressures to $\sim40^{\circ}$ at higher pressures the distance between the laser and the capillary exit is a sensitive parameter, especially when high sample densities are required.

%%%%%%%%%%%%%%%%%%%%%%%%%%%%%%%%%%%%%%%%%%%%%%%%%%%%%%%%%%%%%%%%%%%%%%%%%%%%%%%%%%%%%%%%%%%%%%%%%%%%%%%%%%%%%%%%%%%%%%%%%%%%%%%%%%%%
\subsection{Time of flight spectrometer}

The 1\,kHz repetition rate of the laser is ideal for a time of flight electron/ion spectrometer. With typical flight times in the microsecond range data acquisition can easily be achieved. Whole traces with multiple hits can be digitized and transferred to the computer memory before the following laser shot.

In gas phase photoelectron experiments with limited photon flux, especially in pump-probe experiments, it is desirable to maximize the collection efficiency of the electron spectrometer. While more sophisticated methods such as velocity map imaging~\cite{Eppink1997} or cold target recoil ion momentum spectroscopy~\cite{Doerner2000,Gagnon2008} reveal additional information about the velocity vector of electrons and/or ion fragments, we are primarily interested in the kinetic energy of photoelectrons. 
Up to 4$\pi$\,sr collection efficiency without reduction of the energy resolution can  be achieved with a magnetic bottle electron spectrometer. Based on an electron time of flight spectrometer, it exploits a combination of an inhomogeneous and a homogeneous magnetic field to guide photoelectrons from the ionization region to the detector~\cite{Eland2003,TSUBOI1988,KRUIT1983}. We use a relatively short flight tube length of 460\,mm, similar to a setup that was recently presented and well characterized~\cite{Mucke2012}.
The inhomogeneous magnetic field is produced by a stack of permanent magnets mounted on a xyz manipulator, providing more flexibility compared to electromagnets. 
We use five sintered neodymium disk magnets (25.4\,mm diameter, each 6.4\,mm height, purchased from Grainger Industrial Supplies) that are attached by their magnetic action to a soft iron cylinder (25.4\,mm diameter, 42\,mm height, low carbon magnetic iron, ASTM designation A848-01, purchased from ESPI Metals). On the side of the magnets facing the flight tube a soft iron truncated cone (2.8\,mm top diameter, 12.2\,mm bottom diameter, 4.2\,mm height) is attached to the magnets in order to increase the magnetic flux density $B_z$ in the ionization region. Out of several different cone geometries this one gave the highest $B_z$ values at close distances. With a Hall probe we measure $B_{\text{z}}=640$\,mT at the cone surface and $B_{\text{z,i}}=430$\,mT at 2\,mm distance, approximately the location of the laser beam.
A repeller electrode (25 x 25\,mm$^2$, copper) is located on top of the cone which can be set to small negative voltages (typically $\sim-2$\,V) to ensure the collection of the slowest electrons. For ion detection we apply a positive voltage (typically 2--3\,kV) to this electrode to push the ion fragments towards the flight tube. With the application of pulsed potentials, this configuration allows electron-ion coincidence detection~\cite{Matsuda2011}.
The homogeneous magnetic field is provided by a solenoid attached to the flight tube inside vacuum. With 260 windings and currents of typically 0.7\,A we achieve $B_{\text{z,f}}=0.8$\,mT inside the flight tube. 
The whole interaction region as well as the flight tube are shielded with Mu-metal.

At the end of the drift tube the electrons/ions are detected with a Chevron type dual stage multi channel plate (MCP) detector with a conical anode (40\,mm diameter, Jordan TOF Products Inc.). 
To increase the detection efficiency of slow electrons and to be able to switch from electron to ion detection without changing the detector voltages we use a positive potential of +200\,V at the MCP front surface which is shielded towards the drift tube by a grounded mesh. MCP back and anode are at +1.8\,kV and +2.3\,kV, respectively.
For ion detection we apply a high positive voltage of typically 2-3\,kV to the repeller electrode on the magnet to push ions into the spectrometer. The magnetic field has no meaning in the ion spectroscopy mode due to the large mass/slow velocity of the ions compared to the electrons. To suppress major field distortions by the oven tip a potential can be applied to the oven, which strongly depends on the relative positions of repeller electrode, oven tip and laser focus.
Additionally, voltages can be applied to the flight tube and to a 10\,mm diameter entrance aperture, respectively, to retard the electrons for energy resolution increase or suppression of slow electrons. 

The intrinsic limit of energy resolution is determined by the ratios of the magnetic fields $dE/E = B_{\text{z,f}}/B_{\text{z,i}}\approx 0.2$\%~\cite{TSUBOI1988,KRUIT1983}. The field ratio also dictates the magnification of the magnetic bottle configuration~\cite{TSUBOI1988,KRUIT1983}, in other words, the acceptance area (perpendicular to the magnetic bottle center line) from which electrons reach the MCP detector. The acceptance area depends crucially on the distance between the laser focus and the magnet tip and we estimate it to be in the range of the beam diameter of about 90\,$\mu$m.

Signal pulses are decoupled from the conical anode of the MCP detector and digitized by a high-speed analog-to-digital converter card (Gage Cobra, Dynamic Signals LLC, 2\,GS/s repetition rate, 500\,MHz bandwidth, 8 bit resolution). After transfer to the computer a constant fraction peak-finding algorithm is used to determine flight times.

%%%%%%%%%%%%%%%%%%%%%%%%%%%%%%%%%%%%%%%%%%%%%%%%%%%%%%%%%%%%%%%%%%%%%%%%%%%%%%%%%%%%%%%%%%%%%%%%%%%%%%%%%%%%%%%%%%%%%%%%%%%%%%%%%%%%
\subsection{Calibration with Xe}

\begin{figure}[h]
\centering
\includegraphics[width=85 mm]{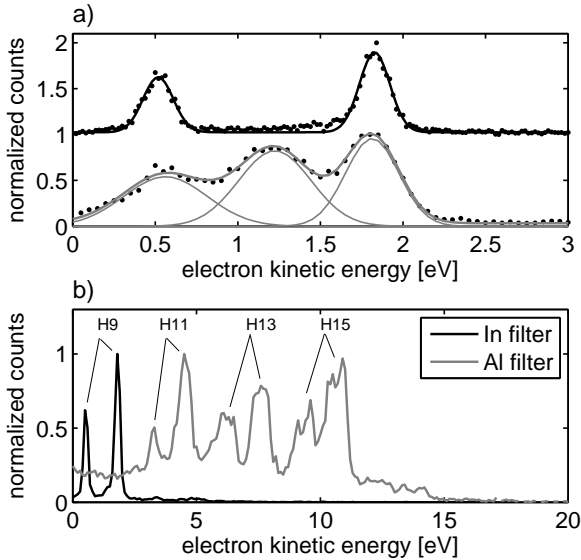}
\caption{Xe reference spactra: (a) PE spectrum of Xe atoms obtained with H9 using the In filter. The upper spectrum (offset by 1) was recorded after optimization of the HHG process to obtain narrow PE lines. The two PE peaks at 0.5\,eV and 1.8\,eV correspond to the Xe 5p$_{1/2}$ and 5p$_{3/2}$ valence electrons. The solid line is a fit of two Gaussian functions to the data points. For the lower spectrum the HHG process was optimized for maximum overall HHG intensity. 
(b) Xe PE spectrum obtained with the In filter (black curve) and with the Al filter (gray curve). The HHG phase matching was optimized in both cases with the In filter.}
\label{fig:Xe}
\end{figure}

We use Xe to calibrate the electron kinetic energy of the TOF spectrometer, optimize the HHG conversion efficiency and determine the H9 photon flux. Figure~\ref{fig:Xe} shows Xe photoelectron spectra obtained with the In filter (H9) and with the Al filter. 
With a H9 photon energy of 14\,eV and binding energies of the Xe 5p$_{1/2}$ and 5p$_{3/2}$ valence electrons of 13.4\,eV and 12.1\,eV, respectively~\cite{Briggs1977}, the photoelctron peaks appear at 0.5\,eV and 1.8\,eV. 
We find that the spectral purity of H9 strongly depends on the HHG phase matching conditions. Maximizing the overall yield of high-order harmonics by maximizing the photocurrent from a gold surface seems to result in poor spectral purity of H9 (lower trace in Fig.~\ref{fig:Xe}a). Instead of two isolated peaks the Xe PE spectrum is composed of three or four peaks. Spectral shifts and splitting of high-order harmonics depending on HHG phase matching conditions have been documented in literature~\cite{Heyl2011,Farrell2009}. 
In contrast, optimization of the HHG parameters by monitoring the shape of the Xe PE spectrum yields the expected double peak structure (upper trace in Fig.~\ref{fig:Xe}a).
A strong variation of the spectral purity of harmonics 11 to 15 can be seen in the lower image of Fig.~\ref{fig:Xe}b. 
The HHG phase matching in this case was optimized for H9 (black curve) and not changed when switching from the In to the Al filter. A successive increasing spectral distortion of harmonics with increasing harmonic number is immediately evident (gray curve).
We attribute the cut-off after H15 to a poor reflectivity of the MgF$_2$ coated Al mirror above H15. 
We applied a small repelling potential to record the Xe PE which we accounted for by subtraction of 1.7\,eV from the actually measured spectrum to match the peak positions to the literature values.

Both PE peaks in Fig.~\ref{fig:Xe}a have the same FWHM of $220\pm10$\,meV. We estimate the energy resolution of our magnetic bottle spectrometer to be $\Delta E/E\approx4$\%, based on documented literature values ($\sim3.3$\% and $\sim1.6$\% for spectrometers with  0.6\,m and 2.4\,m length, respectively ~\cite{Mucke2012,Lablanquie2007}). The absolute energy resolution is $\Delta E=110\pm30$\,meV  (averaged over the two peaks, which appear at 2.2\,eV and 3.5\,eV, respectively). We thus obtain a spectral width of H9 of $190\pm60$\,meV.

We estimate the photon flux of H9 based on PE count rates obtained from Xe background gas. For a fundamental pulse energy of 3.5\,mJ we obtain $\sim1\cdot10^{7}$\,photons/pulse in the TOF ionization region, corresponding to $\sim2\cdot10^{8}$\,photons/pulse in H9 from the high-order harmonic generation process. This estimate is obtained from the sum of Xe 5p$_{1/2}$ and 5p$_{3/2}$ photoelectron counts per pulse (c.f., Fig.~\ref{fig:Xe}), a photoionization cross section of 64.2\,Mb at 14\,eV~\cite{Samson2002}, an estimated overall electron detection efficiency of 0.5~\cite{Mucke2012,Lablanquie2007} and an estimated magnetic bottle acceptance area diameter of 180\,$\mu$m (twice the beam FWHM diameter). 
We thus obtain a H9 conversion efficiency of $2\cdot10^{-7}$, which can be compared to literature values of $\sim2\cdot10^{-6}$ for lower-order harmonics in Ar~\cite{Brabec2000}.

%%%%%%%%%%%%%%%%%%%%%%%%%%%%%%%%%%%%%%%%%%%%%%%%%%%%%%%%%%%%%%%%%%%%%%%%%%%%%%%%%%%%%%%%%%%%%%%%%%%%%%%%%%%%%%%%%%%%%%%%%%%%%%%%%%%%
%%%%%%%%%%%%%%%%%%%%%%%%%%%%%%%%%%%%%%%%%%%%%%%%%%%%%%%%%%%%%%%%%%%%%%%%%%%%%%%%%%%%%%%%%%%%%%%%%%%%%%%%%%%%%%%%%%%%%%%%%%%%%%%%%%%%
\section{Results and Discussion}

To demonstrate the versatility of our spectrometer we show time-resolved pump--probe photoion and photoelectron spectra where ionization is either achieved by a multi-photon 800\,nm transition or by a single 14\,eV photon (H9, 89\,nm). As a first sample we investigate the polycyclic aromatic hydrocarbon perylene (C$_{20}$H$_{12}$, mass = 252\,u). Perylene is a strong absorber at 400\,nm (S$_1\leftarrow \text{S}_0$ excitation)~\cite{Nichols2013} with a gas phase excited state lifetime of 5\,ns~\cite{Fourmann1985} and negligible Stokes-shift~\cite{Nichols2013}. 
The fluorescence yield of the S$_1$ state in the gas phase is 97\%~\cite{Sonnenschein1984}.

%%%%%%%%%%%%%%%%%%%%%%%%%%%%%%%%%%%%%%%%%%%%%%%%%%%%%%%%%%%%%%%%%%%%%%%%%%%%%%%%%%%%%%%%%%%%%%%%%%%%%%%%%%%%%%%%%%%%%%%%%%%%%%%%%%%%
\subsection{800\,nm multi-photon probe: photoion detection}

\begin{figure}[h]
\centering
\includegraphics[width=85 mm]{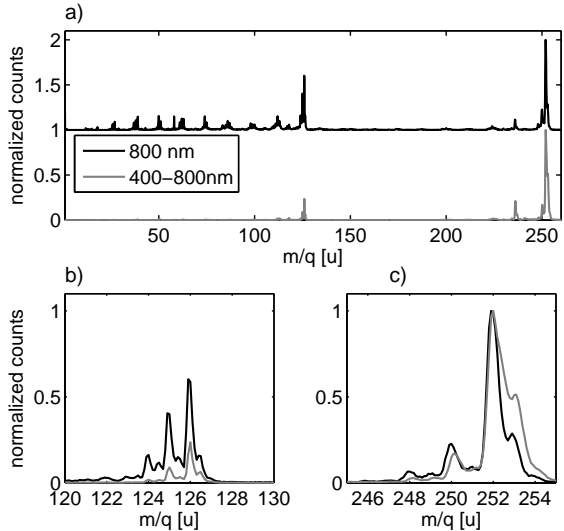}
\caption{(a) Photoion spectra obtained with solely 800\,nm multi-photon ionization (black curve, offset by 1, $1.7\times10^{13}$\,W/cm$^2$) and 400--800\,nm pump--probe photoionization at positive time delay (gray curve, 800\,nm: $5.1\times10^{12}$\,W/cm$^2$). The lower images show regions of the doubly charged parent at 126\,u (b) and the parent at 252\,u (c). 
}
\label{fig:400800ions1}
\end{figure}

Figure~\ref{fig:400800ions1} shows a mass spectrum of perylene obtained with 800\,nm photoionization (black curve) and 400--800\,nm pump--probe photoionization at positive time delay (the 400\,nm pump pulse arrives before the 800\,nm probe pulse, gray curve). Strong field multi-photon ionization from the S$_0$ ground state with 800\,nm at $1.7\times10^{13}$\,W/cm$^2$ leads to significant fragmentation and multiple ionization, as expected by comparison to the ionization of similar size polycyclic aromatic hydrocarbons in this laser intensity range~\cite{Markevitch2004}. The regions around the parent ion peak at 252\,u (Fig.~\ref{fig:400800ions1}b) and the doubly charged parent ion peak at 126\,u (Fig.~\ref{fig:400800ions1}c) are shown in more detail. 
We estimate the mass resolution of our spectrometer in ion mode to be at least 200 for a repeller voltage of $+3$\,kV. We note that the ionization volume is very small due to a tight focusing of the 800\,nm beam ($\sim90$\,$\mu$m FWHM) and the non-linearity of multi-photon ionization process. 
Interestingly, for both perylene$^+$ and perylene$^{++}$ detachment of H$_2$ dominates over single H atom detachment. The peaks at 253\,u and 126.5\,u result from the 1.1\% natural abundance of the $^{13}$C isotope. 

\begin{figure}[h]
\centering
\includegraphics[width=65 mm]{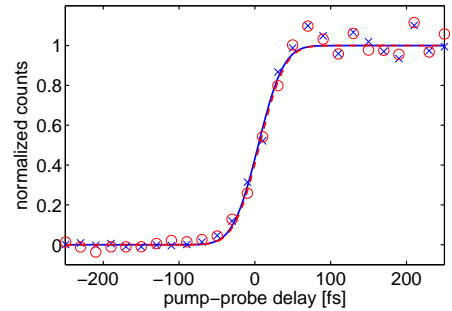}
\caption{Time-dependent 400\,nm--800\,nm pump--probe photoion signal measured at 126\,u (blue crosses and blue solid line) and at 250\,u (red circles and red dashed line). The rising edge is fitted with an error function.  }
\label{fig:400800ions2}
\end{figure}

The 400\,nm--800\,nm pump--probe spectrum is shown in gray in Fig.~\ref{fig:400800ions1}. We chose the laser intensities such that we obtain a negligible signal from the pump or probe laser alone ($\sim50$ counts per second, cps) as compared to the pump--probe signal ($\sim5$\,kcps). The perylene$^+$ signal at 252\,u seems to be saturated, indicated by an unrealistically high signal at 253\,u. 
The time-dependent molecular response as function of the pump--probe delay is shown in Fig.~\ref{fig:400800ions2}. Plotted are the time-dependent ion signals of C$_{20}$H$_{10}^+$ (perylene minus H$_2$ ion at 250\,u) in red (circles and dashed line)  and the doubly charged parent ion perylene$^{++}$ (126\,u) in blue (crosses and solid line). The duration of the signal increase around time zero is proportional to the cross correlation of the temporal profiles of the 400\,nm pump and the 800\,nm probe pulse. Assuming that pump and probe pulses have a Gaussian temporal profile, we fit the data points with the cumulative distribution function of a Gaussian normal distribution 

\begin{equation}
	y=1/2\left[1+\text{erf}\left(\frac{x-\mu}{\sqrt{2\sigma^2}}\right)\right],
	\label{eq:erf}
\end{equation}

where erf() is the errorfunction, $\mu$ is the mean and $\sigma$ is the standard deviation. We obtain the FWHM of the cross-correlation as $\Delta T=2\sqrt{2\text{ln}(2)}\cdot\sigma=62\pm6$\,fs (averaged over both masses). 
To determine the 400\,nm probe pulse duration we perform an autocorrelation measurement detecting perylene ions, which reveals a FWHM of $98\pm6$\,fs. Ionization of ground-state perylene with 400\,nm light is a process of order 1.6, which results from the intensity dependence of the ion signal and indicates that intermediate resonances contribute to the multi-photon process. By calculating the autocorrelation integral for Gaussian time profiles we deduce a 400\,nm pulse duration of $58\pm4$\,fs. Comparison with the 400--800\,nm cross-correlation shows that the limiting factor comes from the 400\,nm pulse duration.

%%%%%%%%%%%%%%%%%%%%%%%%%%%%%%%%%%%%%%%%%%%%%%%%%%%%%%%%%%%%%%%%%%%%%%%%%%%%%%%%%%%%%%%%%%%%%%%%%%%%%%%%%%%%%%%%%%%%%%%%%%%%%%%%%%%%
\subsection{H9 single-photon probe: photoelectron detection}

\begin{figure}[h]
\centering
\includegraphics[width=85 mm]{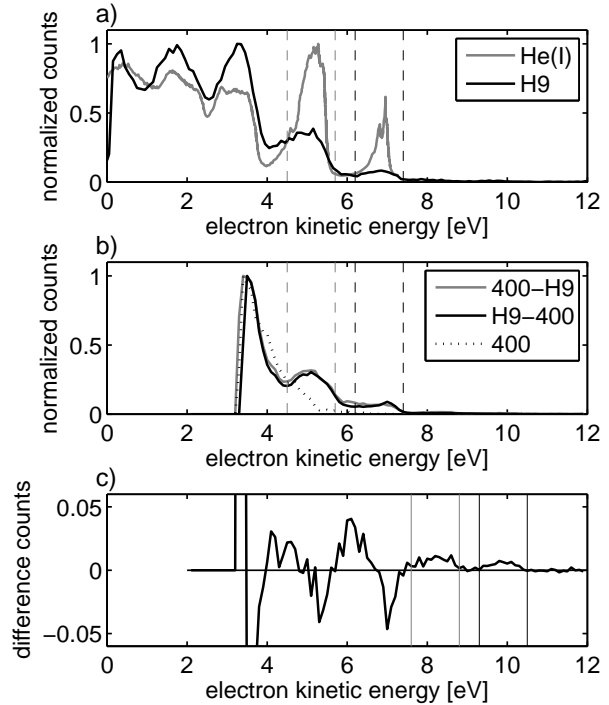}
\caption{(a) Perylene PE spectrum obtained with H9 (14\,eV, black curve) compared to a reference spectrum (gray curve, shifted in energy to be comparable to our result)~\cite{Boschi1972}. 
(b) Pump--probe signal (300\,fs time delay, gray curve), probe--pump signal ($-700$\,fs time delay, black solid curve) and 400\,nm pump-only (black dotted curve) spectra. 
(c) Difference spectrum obtained as difference of the pump--probe signal and probe--pump signal shown in (b). 
 }
\label{fig:400H9electrons1}
\end{figure}

The photoelectron spectrum of perylene observed with H9 (14\,eV) (Fig.~\ref{fig:400H9electrons1}a, black curve) agrees well with a reference spectrum obtained with a He(I) emission lamp at 21.6\,eV~\cite{Boschi1972}. 
The vertical ionization potential of perylene is 7\,eV and the spectrum consists of five bands in the energy range of 7\,eV to 14\,eV~\cite{Boschi1972}. The observed transitions could be assigned to cationic states based on time-depended density functional theory and calculated vertical excitation energies agree within $\sim0.1$\,eV with the experimental values~\cite{Hirata1999,Halasinski2003}. The photoelectron spectrum agrees well with the cationic absorption bands observed in matrix isolation~\cite{Szczepanski1993}.

%While the peak positions and widths are well reproduced the peaks at higher kinetic energy appear to be weaker with H9 than in the reference spectrum. 

We obtain 400\,nm--H9 pump--probe difference spectra by subtracting the probe--pump (negative time delay) signal from the pump--probe (positive time delay) signal, as shown in Fig.~\ref{fig:400H9electrons1}b and c.
A careful choice of the pump pulse intensity is very important  in order to generate sufficient population in the S$_1$ state and at the same time limit multi-photon excitation by the pump pulse. A 400\,nm pulse energy of 650\,nJ ($4.4\cdot10^{10}$\,W/cm$^2$) turned out to be a good compromise (we estimate that about 17\% of the molecules are excited, assuming a 64\,Mb excitation cross section \cite{Dixon2005a,Berlman1971}).
At these laser intensities the pump pulse alone produces a relatively large amount of photoelectrons (Fig.~\ref{fig:400H9electrons1}b, black dotted curve), probably because of resonant enhancement of the two photon transition from S$_1$ into the continuum. To decrease the load of our MCP detectors we apply a negative potential to the drift tube, which we account for in the spectra. 
The retardation is reflected in the PE spectra by a cut at 3.2\,eV.
Fig.~\ref{fig:400H9electrons1}b compares a 400\,nm--H9 pump--probe spectrum (300\,fs time delay, gray curve) with a H9--400\,nm probe--pump spectrum ($-700$\,fs time delay, black curve). The corresponding pump--probe difference spectrum is shown in Fig.~\ref{fig:400H9electrons1}c.
Two new PE bands in the difference signal appear at 8.2\,eV and 10.1\,eV and can be attributed to ionization of the excited molecule in the S$_1$ state to the D0 and D1 continuum, respectively. The energy of the emerging bands is consistent with excitation by the 3\,eV (400\,nm) pump pulse.
Consequently, a transfer of population from S$_0$ to S$_1$ reduces the PE signal originating from S$_0$ resulting in a negative difference signal.
This is directly evident for the two highest energy peaks at 5.1\,eV and 7.0\,eV, which shift to 8.2\,eV and 10.1\,eV, respectively (corresponding energy regions are marked with vertical dashed and solid lines in Fig.~\ref{fig:400H9electrons1}). 
The difference signal maximum at 6.1\,eV corresponds to a shift of the peak at 3.3\,eV. 
In the energy range between 5\,eV and 7.5\,eV positive contributions (population of S$_1$) and negative contributions (depletion of S$_0$) to the difference signal overlap, rendering the determination of band shifts complicated.
Below 5\,eV the difference spectrum becomes very noisy due to the strongly increasing 400\,nm electron signal.

\begin{figure}[h]
\centering
\includegraphics[width=65 mm]{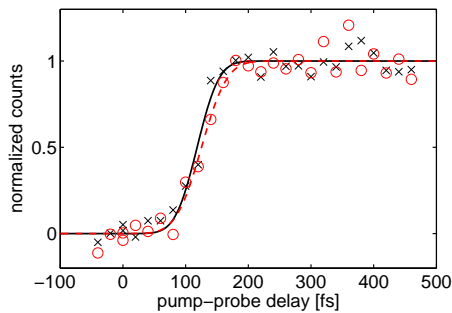}
\caption{Time-dependent rise of the transient peaks in the difference spectrum at 8.2\,eV (red circles and dashed red line) and 10.1\,eV (black circles and solid black line, c.f., Fig.\ref{fig:400H9electrons1}c).  }
\label{fig:400H9electrons2}
\end{figure}

To obtain the time-dependent PE signal we integrate the difference signal in the energy regions of the shifted PE lines (marked with vertical lines in Fig.~\ref{fig:400H9electrons1} c) for different pump--probe delays. The transient PE signals are shown in Fig.~\ref{fig:400H9electrons2}. 
We use a random delay sequence to avoid systematic errors and the data points are more closely spaced around time zero. By fitting equation~\ref{eq:erf} to the data points we obtain a cross correlation FWHM of $68\pm10$\,fs (averaged over both PE energy regions). With a 400\,nm pulse duration of $58\pm4$\,fs (see above) we estimate the pulse duration of H9 to be $35\pm8$\,fs.

%%%%%%%%%%%%%%%%%%%%%%%%%%%%%%%%%%%%%%%%%%%%%%%%%%%%%%%%%%%%%%%%%%%%%%%%%%%%%%%%%%%%%%%%%%%%%%%%%%%%%%%%%%%%%%%%%%%%%%%%%%%%%%%%%%%%
%%%%%%%%%%%%%%%%%%%%%%%%%%%%%%%%%%%%%%%%%%%%%%%%%%%%%%%%%%%%%%%%%%%%%%%%%%%%%%%%%%%%%%%%%%%%%%%%%%%%%%%%%%%%%%%%%%%%%%%%%%%%%%%%%%%%
\section{Conclusion}

We have described a time-resolved pump-probe photoelectron and photoion spectrometer to study photoinduced relaxation processes in gas-phase molecules. Ultrashort pulses in the VUV region obtained from high-order harmonic generation are used as probe.
A simple and cost efficient way has been presented to isolate the $9^{\text{th}}$ harmonic of a 800\,nm fundamental from the harmonic spectrum by means of an In metal filter in combination with a normal incidence MgF$_2$ coated Al mirror. 
The probe photon energy of 14\,eV allows to address the whole set of valence electron levels in many molecules.
Due to the fact that the probe pulse also ionizes the ground state, difference spectra between reversed time delays are used to infer molecular transients. 
The H9 pulse duration of $35\pm8$\,fs has been obtained from cross correlation measurements on perylene molecules. This is longer than expected, most probably due to a fundamental pulse chirp that serves to optimize the spectral bandwidth and intensity of the VUV probe pulse.

In future experiments, electronic relaxation in molecules with non-adiabatic transitions can be followed all the way to the (vibrationally hot) electronic ground state. 
The high photon energy allows mapping of excited states onto a larger domain of the continuum, as compared to crystal-based harmonic probes. This increases the ability to disentangle electronic from vibrational dynamics for internal conversion between neutral excited states with different Koopmans-type correlations to ionic states.
In addition to VUV single-photon probe, 800\,nm multi-photon probe can easily be achieved with our setup, without influencing the beam geometry or pump-probe delay. In multi-photon ionization, which is often used in time-resolved studies, neutral excited molecular states might interfere with the probe scheme and thereby obscure the genuine relaxation dynamics. Direct comparison of the two probe schemes will reveal the influence of such intermediate resonances and thereby provide valuable information about excited state dynamics during the relaxation process.

%%%%%%%%%%%%%%%%%%%%%%%%%%%%%%%%%%%%%%%%%%%%%%%%%%%%%%%%%%%%%%%%%%%%%%%%%%%%%%%%%%%%%%%%%%%%%%%%%%%%%%%%%%%%%%%%%%%%%%%%%%%%%%%%%%%%
\subsection{Acknowledgements}

This work was supported by the AMOS program within the Chemical Sciences, Geosciences, and Biosciences Division of the Office of Basic Energy Sciences, Office of Science, U.S. Department of Energy. M. G. acknowledges funding via the Office of Science Early Career Research Program through the Office of Basic Energy Sciences, U.S. Department of Energy. 
M.K. would like to acknowledge support by the Austrian Science Fund  (FWF, Erwin Schr\"odinger Fellowship, J 3299-N20).
J. G. acknowledges funding from the European Research Agency via the FP-7 PEOPLE Program (Marie Curie Action 298210).
T. J. A. W. thanks the German National Academy of Sciences Leopoldina for a fellowship (LPDS2013-14).

%\bibliography{_MarkusKochAll}

%

\end{document}